\documentstyle[prl,aps,preprint]{revtex}
\large

\begin{document}
\draft

\title{Strong-coupling behaviour in discrete \\
Kardar-Parisi-Zhang equations} 
\author{T. J. Newman and A. J. Bray} 
\address{Department of Physics,\\
University of Manchester,\\
Manchester, M13 9PL, UK}
\maketitle
\begin{abstract}
We present a systematic discretization scheme for the
Kardar-Parisi-Zhang (KPZ) equation, which correctly
captures the strong-coupling properties of the
continuum model. In particular we show that the
scheme contains no finite-time singularities in
contrast to conventional schemes.
The implications of these results to i) previous
numerical integration of the KPZ equation, and ii)
the non-trivial diversity of universality classes for
discrete models of `KPZ-type' are examined. The new scheme
makes the strong-coupling physics of the KPZ equation 
more transparent than the original continuum version and allows
the possibility of building new continuum models which may be 
easier to analyse in the strong-coupling regime.
\end{abstract}
\vspace{5mm}

\newpage

\section{Introduction}

Within the field of non-equilibrium interface growth, much attention
has been given to a continuum Langevin description 
first suggested by Kardar, Parisi and Zhang (KPZ)\cite{kpz,rev1}. 
The KPZ equation for the interface profile $h({\bf x},t)$ takes 
the form
\begin{equation}
\label{kpz}
\partial_{t} h = \nu \nabla ^{2}h + \lambda (\nabla h)^{2} + \eta,
\end{equation}
where $\eta ({\bf x},t)$ is a gaussian white noise with zero mean.
One is interested in asymptotic scaling behaviour of this model.
In particular the steady-state two-point correlation function 
$C(|{\bf r}-{\bf r'}|,t-t') \equiv \langle (h({\bf r},t)-h({\bf r'},t'))^{2} 
\rangle $ is expected to assume a scaling form
\begin{equation}
\label{scaling}
C(R,\tau) \sim R^{2\chi} f(R^{z}/\tau),
\end{equation}
where $\chi$ and $z$ are the roughness and dynamic exponents respectively.
There exists a scaling relation between these exponents of the form
$\chi + z =2$.
The determination of these exponents along with the scaling function $f(y)$ are
the prime goals in this field. For substrate dimension $d \le 2$ the exponents
take on non-trivial values for all $\lambda \ne 0$. For $d>2$ these exists
a phase transition between smooth and rough phases. For small $\lambda $
the interface is smooth $(\chi =0)$ - this behaviour is described by 
the so-called weak-coupling fixed-point. For large $\lambda $ the interface
is asymptotically rough and $\chi $ takes on a non-trivial $d$-dependent
value -- this is the physics of the strong-coupling (SC) fixed point.

It is convenient for us to mention at this point the
interesting connection between the KPZ equation and the physics of
directed polymers in a random medium. This may be realised by making the
Hopf-Cole transformation $h({\bf x},t) = (\nu/\lambda )\ln w({\bf x},t)$
which yields the equation
\begin{equation}
\label{diff}
\partial_{t} w = \nu \nabla^2 w + (\lambda/\nu) \ w \ \eta . 
\end{equation}
The field $w$ is interpreted as the restricted partition function
of the directed polymer (which is more clear if one rewrites the
above equation as a Feynman path integral.)
Although the equation is linear, the noise is now multiplicative which
forbids any simple analytic progress.

In fact there are no systematic methods available to study the SC
behaviour of the KPZ equation analytically. 
The most useful technique to date is a
mode-coupling scheme\cite{mc} which certainly gives non-trivial behaviour
in higher dimensions, but suffers from the draw-back that it
is very difficult to correct systematically. (In particular the mode-coupling
scheme indicates $d=4$ as an upper critical dimension above which
the exponents take on mean-field values; i.e. $\chi =0$.)

In such a situation, numerical methods take on a supreme significance.
In attempting to understand the KPZ model, two main numerical methods
have been employed. The first is that of simulating microscopically
motivated models which are believed on physical grounds to lie within
the KPZ universality class. Such models include Eden growth\cite{eden},
polynuclear growth models\cite{png}, the hypercube stacking model\cite{hcs}
and the Kim-Kosterlitz model\cite{kk}.
The dynamical exponents measured for these different models all lie
within some small range, although numerical precision is not good enough
(especially in higher dimensions) to give a definitive answer as to whether
these models share the same universal features. The more subtle question
of whether their universal features are actually those of the KPZ equation,
relies upon measuring the exponents (and in principle the scaling functions)
for the KPZ equation itself. The lack of analytic techniques means that
such measurements may only be made by direct numerical integration of
a discretized form of the KPZ equation. (Alternatively one may perform
numerical studies of the directed polymer analogue via transfer matrix methods.)
Such a numerical procedure has
been carried out by several groups over the years 
\cite{af,grant,wolf}
with varying degrees of success.
A common feature of these attempts is that numerical integration becomes
increasingly difficult as the value of the coupling constant is raised.
(The instability for large coupling was also noted in ref.\cite{kard}.)
This is unfortunate as ideally one would like to be deep inside the
SC region, which for higher dimensions implies that the dimensionless 
coupling constant is very large. In this paper we shall show that the choice
of spatial discretization of the KPZ equation is crucial in order to ensure
stability under integration, and more importantly to ensure that the
discrete model is still within the universality class of the original
continuum model. The discretization schemes used in previous studies
unfortunately fail on both these counts and the results so obtained must
be viewed with scepticism. 

The remainder of this paper takes the following form. In the next section
we study in some detail the conventional discretization schemes used
to mimic the continuum KPZ equation. We shall show that they contain
defects such as instabilities and ghost fixed points which are absent in
the continuum. In section 3 we motivate a new discretization scheme which
is guaranteed to be stable and which also contains no unphysical fixed points.
This model is the ideal basis for a fresh numerical investigation of
the KPZ equation. In section 4 we look at the SC properties of this discrete
model and find that it is physically more transparent than its continuum
counterpart. Using this new insight we make some speculative propositions
concerning new continuum models which
are intrinsically SC versions of KPZ. The hope is that these models are
more analytically susceptible than the KPZ equation. We end with section 5,
in which we discuss the nature of universality in the space of discrete
`KPZ-type' models, along with giving our conclusions.

\section{Conventional discretization schemes}

In order to perform numerical integration of the KPZ equation it is
necessary to discretize both space and time. In this paper we shall
be exclusively concerned with the delicate nature of the spatial
discretization. Temporal discretization is a less challenging problem,
and there exist many techniques (such as predictor-corrector methods)
which help to stabilize the integration. So henceforth we will keep
time continuous. Now the standard spatial discretization
for the KPZ equation is performed as follows. One replaces the function
$h({\bf x},t)$ by a set of fields $h_{i}(t)$ where the subscript is a
lattice index. The lattice is taken to be hypercubic with spacing $a$.
One then rewrites Eq.(\ref{kpz}) as
\begin{equation}
\label{disc}
d h_{i}(t)/dt = (\nu/a^{2})\sum \limits _{j} (h_{j}(t)-h_{i}(t))
+ (\lambda /2a^{2}) \sum \limits _{j} (h_{j}(t)-h_{i}(t))^{2} + \eta _{i}(t),
\end{equation}
where the sums are over nearest neighbour sites.

There is a minor subtlety here which we shall address immediately before
continuing. In writing the discrete form of $(\nabla h)^{2}$ there are
two obvious choices. The one favoured in past 
work\cite{af,grant,wolf} 
takes the form
(in $d=1$ for simplicity)
\begin{equation}
\label{grad1}
(\partial h/\partial x)^{2} \rightarrow (1/4a^{2})(h_{i+1}-h_{i-1})^{2} 
\end{equation}
whilst the alternative is
\begin{equation}
\label{grad2}
(\partial h/\partial x)^{2} \rightarrow (1/2a^{2})[(h_{i+1}-h_{i})^{2}+
(h_{i}-h_{i-1})^{2}] 
\end{equation}
At this level of the discussion we state that the second form is preferable,
as the first form has the undesirable {\it lattice} feature that it vanishes 
when the 
height at site $i$ is a symmetric peak. Studying surface morphologies
generated by the first form shows that such sites get `left behind' by
their neighbours, as this site receives no driving force from the
nonlinearity. This point is academic in the light of the remaining discussion
in this section.

We shall take two approaches in order to illustrate the problems with
the above discretization scheme. Both approaches are concerned with
the deterministic version of the problem, and both yield the result
that even in the absence of noise the above discretization scheme is
unstable, which is entirely unphysical as the deterministic continuum
model is completely stable and has a single large time solution of
$h={\rm const.}$ (for bounded intial data.) 
In the first approach we compare the numerically integrated solution
of Eq. (\ref{disc}) with the exact solution of the deterministic continuum
equation, for the simple initial condition of a top hat. In the second
approach we exactly solve the discrete equation for some simple cases
and show explicitly that the solutions are unstable. The inherent
instability becomes progressively worse as the coupling $\lambda $ is
increased in value. 

So, beginning with the first approach we consider the deterministic
version of the KPZ equation. An exact solution of this equation
may be obtained via the Hopf-Cole transformation and takes the form
\begin{equation}
\label{exsol}
h({\bf x},t) = (\nu/\lambda) \ln \left \lbrace \int d^{d}y g({\bf x}-{\bf y},t)
\exp [\lambda h({\bf y},0)/\nu ] \right \rbrace
\end{equation}
where $g({\bf x},t)$ is the heat kernel. 

We consider $d=1$ for simplicity.
Consider the simple initial condition of a top hat of height $H$ and
width $2b$ centred at the origin. The above solution then simplifies to
\begin{equation}
\label{tophat}
h(x,t) = {\nu \over \lambda} \ln \left \lbrace 1 + {\left (
\exp(\lambda H/\nu) - 1 \right ) \over 2} \ \left \lbrack
{\rm erf}\left ( {b-x \over (4\nu t)^{1/2}} \right ) +
{\rm erf}\left ( {b+x \over (4\nu t)^{1/2}} \right ) \right \rbrack
\right \rbrace,
\end{equation}
where ${\rm erf}(z)$ is the error function\cite{as}.
It is easy to convince oneself that $h(x,t)$ is a monotonically
decreasing function of time for $|x| \le b$. In other words, 
the block decays away to zero, for any size of the coupling
constant $\lambda $. If one now uses Eq.(\ref{disc}) to evolve the
top hat function one finds that this smooth decay only occurs when
$\lambda $ is small. In fact the value of the coupling where
unphysical behaviour sets in is of order $\lambda _{c} =\nu/H$. In fig.1 we
plot the evolution of the height at the centre of the top hat and at
one edge as a function of time, for $\lambda = 10\lambda _{c}$
 -- curves corresponding to the exact solution
given in Eq.(\ref{tophat}), and direct numerical integration of
the discrete scheme of Eq.(\ref{disc}) are shown.
One sees that the discrete equation fails completely to capture the
correct evolution. This effect cannot be corrected by reducing the
time step. It is a consequence of the spatial discretization and would
occur even for continuous time.

This result is very disappointing
as $\lambda _{c}$ demarcates the transition between weak and strong
coupling behaviour for this simple problem. From the exact solution 
one sees that interesting and revealing KPZ physics is present. For
$\lambda \ll \lambda_{c}$ the nonlinearity is essentially unimportant
and the block diffuses away. More precisely, the centre of the block
decays as 
\begin{equation}
\label{decay}
h(0,t) \sim {bH \over (\pi \nu t)^{1/2} } .
\end{equation}
This power-law decay also holds for $\lambda \gg \lambda _{c}$ but only
for {\it extremely} large times. When the coupling exceeds its critical
value, a new time scale $t^{*}$ emerges, which takes the form
$t^{*} = (b^{2}/\pi \nu)\exp (2\lambda /\lambda _{c})$.
The block diffuses away only
for $t \gg t^{*}$. For $t < t^{*}$ the
centre of the block is essentially frozen, due to the nonlinearity pushing
out the sides of the block. In fact the decay of the centre in this time regime
is of the form 
\begin{equation}
\label{decay2}
h(0,t) \sim H - (\nu/2\lambda) \ln (\pi \nu t /b^{2}) .
\end{equation}
In Appendix A we discuss in more detail the relevance of this result
to the SC physics of the noisy KPZ equation.

The discrete equation fails for the top hat intial condition due to
the large height deviation which occurs over a few lattice sites.
We have tested the discrete equation on smooth initial conditions
such as an inverted parabola and a gaussian, and it reproduces the
correct evolution. In the stochastic version of the model, the
noise is constantly producing discontinuities into the interface,
so any discrete scheme must be stable in the presence of large,
local height deviations.

So to summarize the result of the first approach, we see that even for a
simple scenario -- that of the deterministic evolution of a block
profile -- the discrete equation fails to describe the strong-coupling
physics.

We now turn to the second approach -- 
that of solving the discrete equation (\ref{disc}) explicitly
(again in the deterministic version) for some simple situations. As a warm-up
we consider the ostensibly artificial problem of just three sites on a ring.
We are led to consider the coupled equations 
\begin{eqnarray}
\label{three}
\nonumber
dh_{1} /dt & = & 
h_{3}+h_{2}-2h_{1}+(\lambda/2)[(h_{1}-h_{3})^{2}+(h_{2}-h_{1})^{2}] \\
dh_{2} /dt & = & 
h_{1}+h_{3}-2h_{2}+(\lambda/2)[(h_{2}-h_{1})^{2}+(h_{3}-h_{2})^{2}] \\
\nonumber
dh_{3} /dt & = & 
h_{2}+h_{1}-2h_{3}+(\lambda/2)[(h_{3}-h_{2})^{2}+(h_{1}-h_{3})^{2}] 
\end{eqnarray}
where we have set $\nu = 1 = a$.
Eliminating the soft mode (the mean height) we have two independent modes
which we combine as $u_{1} = (\lambda/6)(h_{1}-h_{2})$ and 
$u_{2} = (\lambda/6)(h_{2}-h_{3})$.
The above equations now read (after rescaling $t \rightarrow t/3$)
\begin{eqnarray}
\label{threeu}
\nonumber
du_{1}/dt & = & u_{1}(-1+u_{1}+2u_{2}) \\
du_{2}/dt & = & u_{2}(-1-u_{2}-2u_{1})
\end{eqnarray}
The alternative discretization of $(\nabla h)^{2}$ yields the same equations as
above, but with $\lambda \rightarrow -2\lambda$. 
Given this is a deterministic system, we expect on the basis of the
deterministic KPZ equation for the final state to be that of
all three heights equal, independent of the initial condition, i.e.
$u_{i}(t) \rightarrow 0$. That this is not the case for this discrete model
may be seen by finding the fixed points of Eqs. (\ref{threeu}),
along with their associated stability. There are actually four fixed points, 
only one of
which is the physically reasonable one. The remaining three fixed points
correspond to one particular configuration, allowing for the threefold 
translational
degeneracy. In fig.2 we show the dynamic flows
of the above equations and sketch the approximate boundaries of stability.
The insets shows the configurations relating to the various fixed points. 
One sees that the model is only driven to the correct fixed point for
initial conditions which lie within the boundaries of stability. The linear 
dimensions
of the region (in terms of the original $h_{i}$ variables) 
are of order $1/\lambda$. This indicates once again that as $\lambda $ is
increased the stability of the discrete equations decreases.

Although this problem seems far from any interesting situation, this is
not the case. Since we have enforced periodic boundary conditions, the
above result shows that any periodic chain whose length is a multiple of
three will be unstable since the chain only has to partition itself into
groups of three for the unstable fixed points found above to become active.
Of course, as the number of particles increases, there will be an increasingly
large number of fixed points corresponding to different modes of the system.
That there exists at least one such unstable mode (as we have demonstrated 
above) is
sufficient for our purposes. This result states that the discrete scheme
given in Eq.(\ref{disc}) is intrinsically unstable due to the existence
of ghost (i.e. unphysical) fixed points which separate the evolution
of the system from its true asymptotic state. 

To illustrate this point further, we briefly consider one more model solution
of the discrete equations, this time for an arbitrarily large number
of sites. Consider Eq.(\ref{disc}) for $N$ height variables $\lbrace h_{i}(t) 
\rbrace$
coupled together via the scheme given in Eq.(\ref{grad1}). 
(For variety, we use the alternative discretization of $(\nabla h)^{2}$.) 
We take the
ends of the chain to be pinned; i.e. $h_{0}(t) = 0 = h_{N+1}(t)$.
Again, the physics of the continuum problem admits only one asymptotic
solution - that of a flat interface with $h_{i}(t)=0$. The explicit
form of the discrete equations is
\begin{eqnarray}
\label{Nheights}
\nonumber
dh_{1}/dt & = & -2h_{1}+h_{2}+(\lambda/4)h_{2}^{2} \\
dh_{i}/dt & = & h_{i-1}-2h_{i}+h_{i+1} + (\lambda/4)(h_{i+1}-h_{i-1})^{2},
 \ \ \ \ \  2 \le i \le N-1 \\
\nonumber
dh_{N}/dt & = & h_{N-1}-2h_{N}+(\lambda/4)h_{N-1}^{2}
\end{eqnarray} 
The task of finding all the fixed points of the above equations is
extremely difficult. We will content ourselves with studying an
obvious fixed point: making the Ansatz $h_{i}(t) = h(t)$, 
we see that such a solution picks out two fixed points. The first
is the physical one $h(t)=0$, whilst the second is $h(t)=4/\lambda $.
An analysis of the physical fixed point reveals, of
course, its {\it linear} stability. The stability analysis of the
unphysical fixed point follows from writing $h_{i}(t) = 4/\lambda 
+ {\hat h}_{i}(t)$. Keeping linear terms only, we have the 
stability equations
\begin{eqnarray}
\label{stabilityN}
d{\hat h}_{1}/dt & = & -2{\hat h}_{1}+3{\hat h}_{2} \\
d{\hat h}_{i}/dt & = & {\hat h}_{i-1}-2{\hat h}_{i}+{\hat h}_{i+1},
\ \ \ \ \ 2 \le i \le N-1 \\
d{\hat h}_{N}/dt & = & 3{\hat h}_{N-1}-2{\hat h}_{N}
\end{eqnarray}
The stability criterion is now reduced to the following question. Does the
matrix ${\bf M}$ defined by $d{\bf {\hat h}}/dt = {\bf M}{\bf {\hat h}}$
have any positive eigenvalues, and if so, how do these eigenvalues depend on
$N$.
Details of a numerical solution are given in Appendix B.
We find the following result: for $N \le 6$ there exists one positive
eigenvalue, whereas for $N > 6$ there exist two
positive eigenvalues which become degenerate and of order unity, in the 
limit of large $N$. We therefore see that this $N$ body problem is also unstable 
and once
again, the instability occurs for height deviations which become
greater than a value of $O(1/\lambda)$.

We shall leave a full discussion of these results for the final section.
At this point we just reiterate the results of this section. We have studied
the deterministic version of the conventional discretization scheme from a 
number of viewpoints. It has been found to yield both unphysical fixed
points, and associated instability, whenever neighbouring height differences
exceed a critical value of $O(1/\lambda)$.
{\it Thus, it has no connection to the
continuum model in the strong coupling regime}. 
The addition of noise to this scheme only lessens its
range of applicability since (even for weak noise) strong fluctuations
will create critical height differences thus driving the evolution
away from the correct asymptotic form.

\section{New discretization scheme}

The previous section has established the entirely negative
result that conventional discretization schemes of the KPZ equation
are in fact physically unrelated to that equation, with
the degree of physical integrity decreasing as one 
increases the coupling constant.
The important question remains -- how does one discretize the KPZ
equation in such a way as to ensure stability and also to 
retain the important physics?
In this section we provide an answer to this question which turns out
to extremely simple, but also physically revealing.

Before doing so, however, we consider the possibility that the discrete scheme 
displayed in Eq.\ (\ref{disc}) can be regularised by the addition of further 
terms, thereby avoiding any instability. Such an approach has recently been 
developed in the context of an exactly soluble mean-field model \cite{Marsili}. 
In its simplest form, it corresponds to the addition of a term 
$\kappa (\nabla h)^2 \nabla^2 h$ to the right-hand side of Eq.\ (\ref{kpz}), 
followed by a `naive discretisation' as in Eq.\ (\ref{disc}). The resulting 
equation is soluble in the infinite-range limit where each height $h_i$ is 
coupled to every other height and, in the limit of weak noise, exhibits an 
interesting phase transition as a function of the variable 
$g=\lambda/\sqrt{\kappa\nu}$ . The strong-coupling phase $g>g_c$ is 
characterised by a double-peaked distribution for the fluctuations of the 
local heights $h_i$ from the mean height \cite{Marsili}. This implies bumps 
of a definite height in the surface, reminiscent of numerical studies of a 
similar regularised model \cite{NS}. This behaviour is, however, quite 
different from that of the supposedly related directed polymer model, 
leading one to question whether, even after regularisation, the discrete 
and continuum models belong to the same universality class. This is the 
question we now address. 

As we mentioned in the Introduction, there is a model intimately
related to the KPZ equation; namely, the directed polymer in a
random medium. The model representation consists of a linear
diffusion equation with multiplicative noise, as shown in
Eq.(\ref{diff}). The spatial discretization
of this equation is completely under control, as the discretization
only affects the Laplacian term. The strong-coupling physics
enters via the local noise term, and will not be affected by
spatial discretization. The course is clear. We implement the spatial
discretization via the diffusion-like equation and then make the
inverse Hopf-Cole transformation on the lattice in order to obtain
a discrete form of the KPZ equation which is guaranteed to be stable
for arbitrary values of the coupling.

So explicitly we rewrite Eq.(\ref{diff}) as
\begin{equation}
\label{disdiff}
dw_{i}/dt = (\nu/a^{2})\sum \limits _{j} (w_{j}-w_{i})
+ (\lambda /\nu) \ w_{i} \ \eta_{i}.
\end{equation} 
We now make the inverse transformation
\begin{equation}
\label{ihc}
w_{i}(t) = \exp [\lambda h_{i}(t)/\nu ]
\end{equation} 
which yields the new discrete form of the KPZ equation:
\begin{equation}
\label{dkpz}
dh_{i}/dt = (\nu ^{2}/a^{2}\lambda) \sum \limits _{j}
\left \lbrace \exp [(\lambda/\nu )(h_{j}-h_{i})] \ - \ 1\right \rbrace
+ \eta _{i} .
\end{equation}

This equation appears unwieldy, but contains a very important message.
It is useful to define local `velocities' $v_{i,j} = F(h_{i}-h_{j})$ 
in such a way that one may write the discrete KPZ equation in the form
\begin{equation}
\label{vels}
dh_{i}/dt = \sum \limits _{j} v_{i,j} + \eta _{i}.
\end{equation}
One then sees that the functional form of these velocities for the
above discretization scheme takes the form
\begin{equation}
\label{velf}
F(x) = (\nu^{2}/\lambda a^{2}) \ [\exp (-\lambda x/\nu) - 1].
\end{equation}
In contrast, the conventional scheme as shown in Eq.(\ref{disc})
gives a functional relationship of the form
\begin{equation}
\label{velf1}
F(x) = - (\nu /a^{2}) x + (\lambda/2a^{2}) x^{2}. 
\end{equation}

Comparing these two forms indicates that the conventional scheme is
a truncated form of the new scheme to second order in $(\lambda x/\nu)$ which
will be accurate so long as $x \sim (h_{i}-h_{j}) < \nu/\lambda$ which
is consistent with our findings in the previous section: the conventional
scheme becomes unstable when neighbouring height differences occur
which are of order $\nu/\lambda$. 
The physics of the KPZ equation in the SC regime is such that 
large local maxima (i.e. those points
which have large positive local height differences) are {\it not} evolved
by the equation. They experience a weak negative force 
of order $(\nu^{2}/\lambda a^{2})$. 
We see that the conventional schemes fail because they actually boost
such points in the positive direction. This unphysical behaviour will
always be present if one attempts to discretize the KPZ equation using
a finite number of terms in the height differences. The correct discretization
schemes (of which Eq.(\ref{dkpz}) is one example) must be such as to
{\it freeze} points which are local maxima.
This effective freezing was already seen in
the deterministic top hat problem studied in section 2.

To end this section we make some practical remarks. The scheme presented
in Eq.(\ref{dkpz}) is unwieldly and will become more so when one discretizes
time as well. In this case one must not discretize time in the $h$ 
representation, but must inverse Hopf-Cole transform the space-time
discretized equation for $w$. It is clear that one may as well numerically
integrate the equation for $w$ and then perform the statistical analysis
on logarithms of that function. In fact one is essentially simulating
the directed polymer problem. This implies that there is no {\it easy}
way of directly integrating the KPZ equation. The only physically
meaningful way of doing this is to use Eq.(\ref{dkpz}), which is 
tantamount to dealing with directed polymers. In fact, there is one
example in the literature of a direct integration of the discretized
equation for $w$ in $d=2$\cite{beccaria}. The authors were motivated
to integrate this form of the equation due to the reported problems in earlier
work of numerical instability for larger values of the coupling constant.
This motivation has been seen in the previous sections of the current work
to be well justified. Their results for the dynamic exponents agree
well with those found in ref.\cite{wolf}, although they stress the
difficulty of determining whether their results were obtained in
the true asymptotic regime. It is certainly worthwhile to extend this
type of numerical analysis to both larger values of the coupling
constant, and to higher dimensions. 

\section{New continuum analogues}

Aside from the practical concerns of how to integrate the KPZ equation,
the previous sections have shed some light on what the SC physics
of the KPZ equation actually means. What we have seen is that for
large values of the coupling constant, local maxima in the interface
become frozen or pinned. The interesting dynamical effects occur
when we pump such a system with noise. Only the largest positive
deviations will remain pinned for a long time, the smaller ones
being swamped by the local noise fluctuations. In other words,
the rare fluctuations become frozen in and therefore assume a much
greater statistical weight.

The question is how sturdy is this type of SC physics. Referring again
to Eq.(\ref{velf}), we see that the SC physics of the KPZ equation
lies in the fact that $F(x)$ is essentially zero for $x>0$. The
precise form of this function for $x<0$ may not be so important --
this part of the function simply controls the rate at which lower
points try to catch up with their frozen higher neighbours. To test
the hypothesis that the KPZ universality class is determined only
through the special features of $F(x)$ for $x>0$, it is interesting
to consider other models which retain this structure.

One such example is the continuum model defined via
\begin{equation}
\label{noib}
\partial_{t} h = D \nabla^{2} h + \eta,
\end{equation}
where
\begin{equation}
\label{nm}
\nonumber
D = \left \{
\begin{array}{ll}
0 \ ,
\hspace{20mm} & \nabla^{2} h < 0 \\
D_{0} \ ,
\hspace{20mm} & \nabla^{2} h > 0
\end{array}
\right.
\end{equation}
This model is a not-too-distant relative of a stochastically
pumped version of Barenblatt's equation\cite{bar,gold}.

The deterministic evolution of this equation consists of regions
of negative curvature being strongly pinned, whilst regions of
positive curvature are diffusively relaxed. In this sense it is
an inherently SC realization of the KPZ equation.
The relaxational time scales are certainly slower in this model
than in the KPZ equation, but as we argued above, if one believes
that the SC evolution is dominated by frozen surface maxima, then
this difference of time scales may not effect the universal features
of the model. The pinning of regions of negative curvature in the
above model actually makes it a prime continuum candidate for the description
of the RSOS model, in which evolution is only possible within
regions of positive curvature (although one must probably be more
careful with the definition of the noise in this case.)
It would be of interest to 
numerically study this model and ascertain whether it shares
the same scaling features as the KPZ equation. Perhaps it is
even possible to perform some analytic treatment. The nice feature
of this model (as with many others which may be generated by altering the
function $F(x)$ for $x<0$) is that it is intrinsically SC in the KPZ sense. 
We leave this speculative section now and end the paper with our more
concrete conclusions.

\section{Discussions and Conclusions}

We have shown that conventional discretization 
schemes\cite{af,grant,wolf} of the
continuum KPZ equation (such as that shown in Eq.(\ref{disc}))
have two failings which forbid them to remain faithful to the
continuum equation, especially in the SC regime. Firstly they
exhibit ghost fixed points (i.e. certain configurations which
are stationary solutions) which are absent in the continuum;
and secondly, these ghost fixed-points are unstable, so that
the models deviate exponentially fast from the physical
evolution. We believe we have clearly demonstrated these
defects such that the previous results obtained from numerical
integration of these schemes must be viewed as unsound.

We have shown that by discretizing the directed polymer version
of the KPZ equation, one may establish a stable, physically
faithful scheme for integration of the KPZ equation. The main
advantage of this new discretization is not so much its practical worth,
but rather its illumination of the true nature of strong-coupling
behaviour in the KPZ equation. On the basis of this new insight
we have postulated a new criterion for models to be in the continuum KPZ
universality class, and have suggested perhaps the simplest
continuum member of this class, which is related to the Barenblatt equation
with a stochastic source.

A perplexing question remains. Although we now understand how to make
discrete models faithful to the continuum KPZ equation, it is 
disconcerting that the discrete forms such as Eq.(\ref{disc}) are
not in the continuum KPZ universality class, when they seem so reasonable at 
first glance. It opens the possibility of a more tenuous link between
microscopic (lattice) models and the continuum KPZ equation. What we
have seen is that SC physics in the KPZ equation requires local
maxima to be strongly pinned. This kind of physics is transparent in
some microscopic models such as the RSOS model, and polynuclear
growth models. However, other microscopic models which might be
better represented by naive lattice versions of the KPZ equation,
will certainly not be in the true KPZ universality class. The cataloguing
of the universality classes of the KPZ equation, and other 
non-equilibrium interface models, remains as ever an
absorbing challenge for the future.

\vspace{10mm}

TJN is grateful to Michael Swift for useful preliminary
discussions, and acknowledges financial support from the 
Engineering and Physical Science Research Council.
AJB thanks Matteo Marsili for discussions. 

\newpage

\appendix

\section{}

In this appendix we extend the results of the top hat problem 
studied in section 2. We shall connect the existence of
a critical coupling constant to the known behaviour of the 
noisy KPZ problem in the vicinity of the weak coupling fixed
point. First, we generalize the results to arbitrary dimension,
by considering a pill-box in dimension $d$, of radius $b$ and
height $H$. Restricting our attention solely to the centre
of the pill-box we have the exact solution
\begin{equation}
\label{pill-box}
h({\bf 0},t) = (\nu/\lambda) \ln \left \lbrace 1 + 
[\exp (\lambda h/\nu ) - 1] \ I(t) \right \rbrace
\end{equation}
where
\begin{equation}
\label{It}
I(t) = {2\over \Gamma (d/2)} \ \int \limits _{0}^{b/(4\nu t)^{1/2}}
dy \ y^{d-1} \ e^{-y^{2}} ,
\end{equation}
and $\Gamma (z)$ is the Gamma function\cite{as}.
For $\nu t \gg b^{2}$ this expression simplifies to
\begin{equation}
\label{pb}
h({\bf 0},t) \sim (\nu/\lambda) \ln \left \lbrace
1 + {[\exp (\lambda H/\nu) -1]\over \Gamma (1+d/2)}
\ {b^{d} \over (4\pi\nu t)^{d/2}} \right \rbrace .
\end{equation}

We see once again that there exists a critical value for the
dimensionless parameter $(\lambda H/\nu )$. 
For $(\lambda H /\nu) \ll 1$ the above expression may be
expanded out to give the simple diffusion result for large times.
Explicitly one has 
\begin{equation}
\label{simplediff}
h({\bf 0},t) \sim b^{d} H / (4\pi \nu t)^{d/2}.
\end{equation}
However, for $(\lambda H/\nu) \gg 1$ we may only expand the 
logarithm for times larger than $t^{*}$ where
\begin{equation}
\label{crittime}
t^{*} = (b^{2}/\nu ) \ \exp (2\lambda H /d\nu) .
\end{equation}
For $t \gg t^{*}$, the centre of the block once again diffuses
away. However, for $t \ll t^{*}$, the centre is essentially frozen,
and decays only logarithmically slowly. Explicitly one has
\begin{equation}
\label{logdecay}
h({\bf 0},t) \sim H - (d\nu /2\lambda) \ln (\nu t/b^{2}) .
\end{equation}

The message we wish to draw from this result is that for any finite
dimension $d$, there exists a critical value of the coupling constant
of order $\lambda _{c} \sim (\nu/H)$, such that for $\lambda \gg \lambda _{c}$,
a surface deviation which has a scale of $H$ will become essentially frozen,
whilst a surface deviation of a smaller scale will diffuse away, almost
unaware of the non-linearity. If the reader feels uncomfortable with the
discontinuous properties of the top hat function, we note that similar
asymptotic results may be obtained by replacing the top hat over the range
$|x| < b$ by an inverted parabola of central height $H$.

Now consider the full noisy KPZ equation, starting from an initially
flat interface. Over an initial time period the surface fluctuations
will evolve according to the linear (or Edwards-Wilkinson\cite{ew}) model.
Accordingly the root-mean-square fluctuations grow in time as
\begin{equation}
\label{ewfluc}
\nonumber
\langle (\delta h)^{2} \rangle ^{1/2} \sim \left \{
\begin{array}{ll}
t^{(2-d)/4} \ ,
\hspace{20mm} & d < 2 \\
(\ln (t/t_{0}))^{1/2} \ ,
\hspace{20mm} & d=2 \\
{\rm const.} \ ,
\hspace{20mm} & d>2
\end{array}
\right.
\end{equation} 
It is our contention that the SC physics of the KPZ equation will come
into effect only when these rms fluctuations become of the same order
as a critical height scale H, which is set by the condition $\lambda H/\nu 
\sim O(1)$. This idea then leads to a prediction of the crossover time, $t_{c}$
from weak to strong coupling behaviour by equating $\langle (\delta h)^{2} 
\rangle ^{1/2}$
with $H \sim \nu/\lambda $. One finds for $d<2$ that
$t_{c} \sim \lambda ^{-4/(2-d)}$. For $d=2$ one has $t_{c} \sim t_{0} \exp 
[C(\nu^{3}/D\lambda^{2})]$ (where $C$ is some number of order unity.)
For $d>2$ the rms fluctuations saturate to a constant value. One then has a very
simple interpretation of the weak/strong coupling phase transition. For small 
$\lambda $,
the critical height $H$ is large and thus exceeds the saturated rms fluctuations 
-- hence
there will be no possibility for the SC behaviour to set in. Conversely, for 
large
$\lambda $, the critical height $H$ is small and will be quickly swamped by the
weak rms fluctuations, such that the SC behaviour is initiated. 

The particular forms for the weak/strong coupling crossover time found above
for $d \le 2$ have been found in previous studies\cite{tang} by an explicit
integration of the one-loop renormalization group flow equations.
The simple derivation above is useful as it brings out the physical
mechanism behind the flow to SC physics -- i.e. the freezing of
rms fluctuations of order $\nu /\lambda $ and the subsequent
non-linear evolution.  

\section{}

In this appendix, we examine the eigenvalue spectrum of the matrix ${\bf M}$
that appears in the stability analysis of the $N$-site problem in section 2.
Explicitly one has the $N \times N$ matrix
\begin{equation}
\label{matrix}
{\bf M} = 
\left ( \begin{array}{ccccccc}
-2 & 3 & 0 & \cdots & 0 & 0 & 0 \\
1 & -2 & 1 & \cdots & 0 & 0 & 0 \\
0 & 1 & -2 & \cdots & 0 & 0 & 0 \\
\vdots & \vdots & \vdots & \ddots & \vdots & \vdots & \vdots \\
0 & 0 & 0 & \cdots & -2 & 1 & 0 \\
0 & 0 & 0 & \cdots & 1 & -2 & 1 \\
0 & 0 & 0 & \cdots & 0 & 3 & -2 
\end{array} \right )
\end{equation}
The interest lies in whether this matrix has positive eigenvalues (which
signal an instability in the original problem), and if so, whether these
eigenvalues survive in the large-$N$ limit.

We determine the eigenvalues by direct numerical computation using the
{\it Mathematica} program.
The result is that for increasing
$N$, there exist two positive eigenvalues which become degenerate at
a value of 0.1214(1) for large-$N$. A sample of the results for finite
$N$ is presented below:

$$
\arraycolsep=4pt
\begin{array}{|c||*{10}{c|}} \hline
N & 4 & 5 & 6 & 7 & 8 & 10 & 13 & 20 & 25 & 29 \\ \hline
l_{1} & 0.303 & 0.236 & 0.199 & 0.175 & 0.160 & 0.141 & 
0.132 & 0.122 & 0.1215 & 0.1214 \\ \hline
l_{2} & - & - & - & 0.000 & 0.047 & 0.091 & 0.108 &
0.121 & 0.1212 & 0.1213 \\ \hline 
\end{array}.
$$

\newpage

{\bf Figure Captions}

\vspace{20mm}

\noindent
Fig.~1: The time evolution of the centre of the top hat function, and
of one edge of the top hat function, for $\lambda = 10\lambda _{c}$.
The curves represent the exact solution as given in Eq.(\ref{tophat}),
and the points represent numerical integration using the
conventional discretization scheme shown in Eq.(\ref{disc}). 
Parameter values are $\nu=1.0, \ a=1.0, \ b=10.0, \ H=2.0$ and
for the numerical integration $\Delta t = 0.002$. For these
values, $\lambda _{c} = 0.5$.

\vspace{10mm}

\noindent
Fig.~2:  The phase space
of the three-site problem. The dashed lines represent the
domain of stability which has linear dimensions
of $O(1/\lambda )$. The insets indicate the site configurations 
corresponding to the various fixed points.

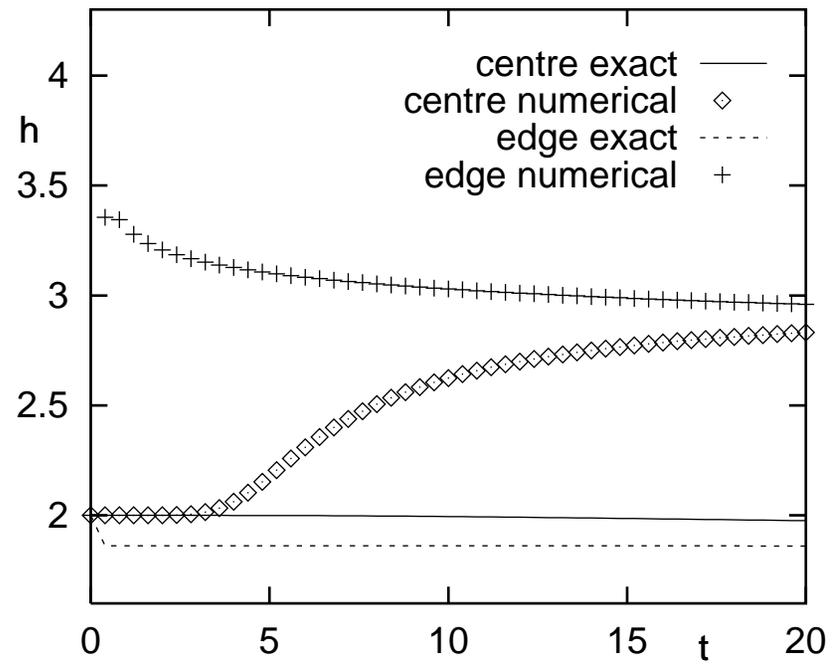
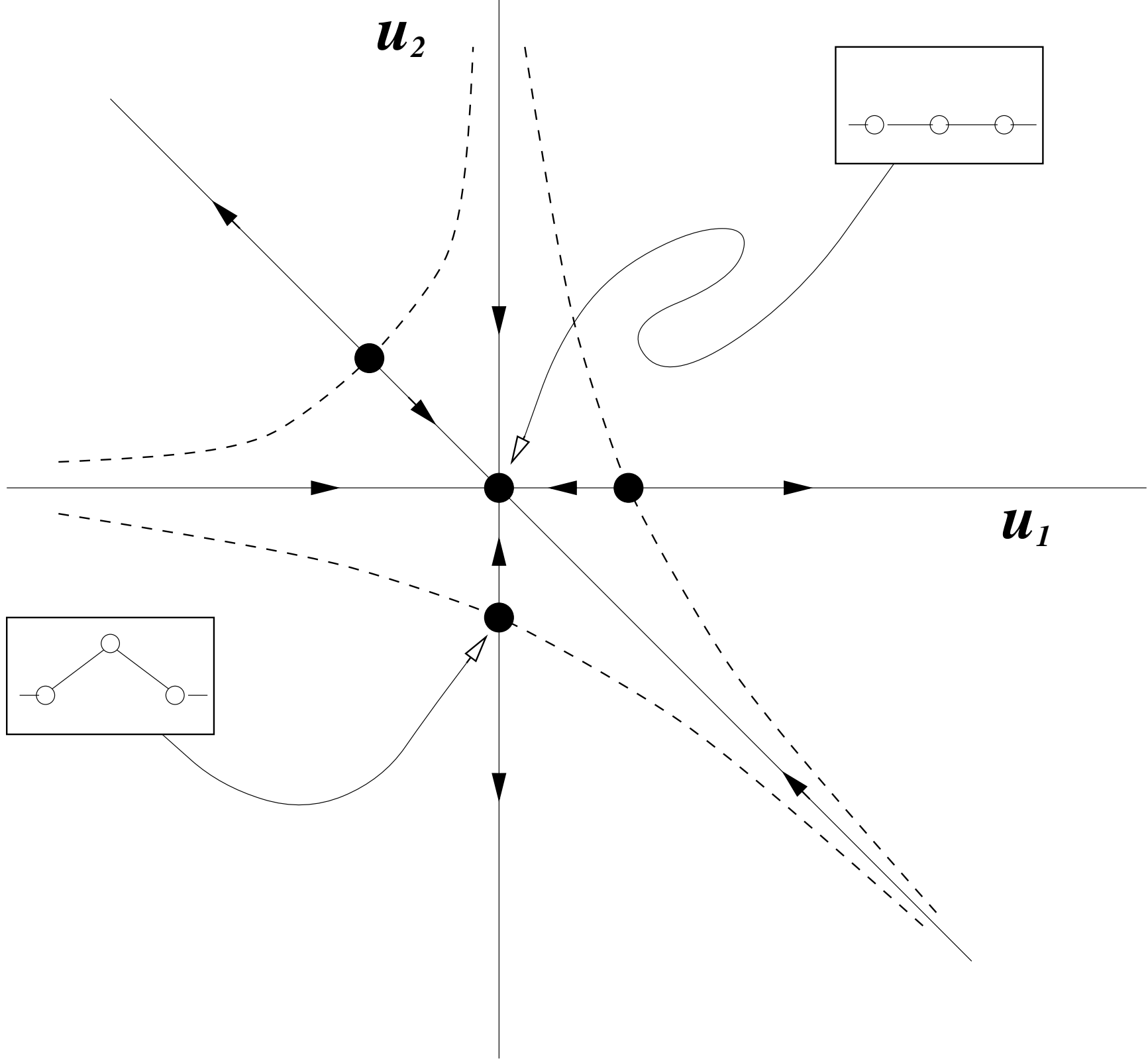
\end{document}